\documentclass[conference]{IEEEtran}
\usepackage{graphicx}
\usepackage{epstopdf}
\usepackage[latin1]{inputenc}
\usepackage{amsmath,amssymb,amscd,latexsym,dsfont}
\usepackage[english]{babel}
\usepackage{tabularx,cite}
\usepackage{graphicx}
\usepackage{algpseudocode}
\usepackage{algorithm}
\usepackage{algorithmicx}
\usepackage{float}
\usepackage{multicol}
\usepackage{psfrag}
\usepackage{comment,psfrag,subfigure,enumerate}
\usepackage{color}
\usepackage{amsthm}

\ifCLASSINFOpdf
\else
\fi
\begin{document}
\IEEEoverridecommandlockouts
\IEEEpubid{\makebox[\columnwidth]{978-1-4799-5863-4/14/\$31.00 \copyright 2014 IEEE \hfill} \hspace{\columnsep}\makebox[\columnwidth]{ }}
\title{Reinforcement-based data transmission in temporally-correlated fading channels: Partial CSIT scenario}

\author{
    \IEEEauthorblockN{Behrooz Makki\IEEEauthorrefmark{1}, Tommy Svensson\IEEEauthorrefmark{1}, Merouane Debbah\IEEEauthorrefmark{2}}
    \IEEEauthorblockA{\IEEEauthorrefmark{1}Department of Signals and Systems, Chalmers University of Technology, Gothenburg, Sweden
    \\\{behrooz.makki, tommy.svensson\}@chalmers.se}
    \IEEEauthorblockA{\IEEEauthorrefmark{2}Alcatel-Lucent Chair - SUPELEC, Gif-sur-Yvette, France
    \\merouane.debbah@supelec.fr}
    \thanks{Behrooz Makki and Tommy Svensson are supported in part by the Swedish Governmental Agency for Innovation Systems (VINNOVA) within the VINN Excellence Center Chase.}
    \thanks{Merouane Debbah has been supported by the ERC Starting Grant 305123 MORE (Advanced Mathematical Tools for Complex Network Engineering).}
}

%Telephone: (800) 555--1212, Fax: (888) 555--1212}


\maketitle
%\onecolumn

\begin{abstract}
Reinforcement algorithms refer to the schemes where the results of the previous trials and a reward-punishment rule are used for parameter setting in the next steps. In this paper, we use the concept of reinforcement algorithms to develop different data transmission models in wireless networks. Considering temporally-correlated fading channels, the results are presented for the cases with partial channel state information at the transmitter (CSIT). As demonstrated, the implementation of reinforcement algorithms improves the performance of communication setups remarkably, with the same feedback load/complexity as in the state-of-the-art schemes.
\end{abstract}
%Then, compared to open-loop communication setups, the implementation of power-adaptive ARQ reduces the average power by ? and ? dB, if a maximum of 2 and 3 retransmissions is utilized, respectively.


% \fi
\IEEEpeerreviewmaketitle
\vspace{-0mm}
\section{Introduction}
In machine learning, reinforcement algorithms refer to the schemes where dynamic parameter adaptation is performed based on a reward-punishment strategy \cite{reinforcement}. The previous trial(s) being successful, more aggressive parameter settings are risked. On the other hand, the parameters of the upcoming trials are designed more conservatively, if the previous gambling fails. Reinforcement learning differs from the standard \emph{supervised} learning in that the correct input/output pairs are never presented, but a reward-punishment signal is used for parameter adaptation, and the goal is to maximize some notion of the cumulative reward. Due to its generality, the reinforcement algorithm is applied in different fields, such as game theory, control theory, simulation-based optimization and statistics, e.g., \cite{1167350,4445757,4581648,6117774}. However, except some works in these last years, e.g., \cite{4581648,6117774,4278411,4305437,4156378,6542770,6503987}, the reinforcement concept has not been well studied in wireless communication.
%there are few works utilizing the reinforcement concepts in wireless communication \cite{4581648,6117774,4278411,4305437,4156378}.

In this paper, we elaborate on the performance of communication systems utilizing reinforcement algorithms. The problem is cast in form of optimizing the data transmission efficiency of wireless networks in the cases with partial channel state information at the transmitter (CSIT). The results are obtained for temporally-correlated fading channels, and the reward-punishment signal is used to dynamically adapt the data transmission rates/powers.

The partial CSIT systems are mainly based on two different channel state information (CSI) feedback models, namely, CSI quantization \cite{33,1715541,ekbatanioutage,Tcomkhodemun,6510028} and hybrid automatic repeat request (HARQ) \cite{6104175,6510028,Tcomkhodemun,outageHARQ,5426285,noisyARQkhodemun,tuninetti2011,greenkhodemun}\footnote{Throughout the paper, we concentrate on the frequency-division duplexing (FDD) communication setups. However, the reinforcement algorithms are applicable in time-division duplexing (TDD) systems as well.}. With CSI quantization, the channel quality information is fed back before the codeword transmission.  The HARQ methods, on the other hand, are based on reporting the decoding status of the previous messages. Here, to emphasize the generally of the reinforcement algorithms, we consider both the quantized CSI and the HARQ feedback models and the results are presented for different metrics. Specifically, considering the point-to-point communication setups, we address the following problems:
\begin{itemize}
  \item\textbf{Problem 1:} \emph{Power-limited throughput maximization in the presence of quantized CSI feedback.} Here, the reward-punishment signal is used for dynamic adaptation of the data transmission rates such that the throughput is maximized. The results are compared with the ones in the static quantization techniques \cite{33,Tcomkhodemun} which, with the same feedback load, show more than $6\%$ throughput increment for a large range of fading correlations.
  \item\textbf{Problem 2:} \emph{Outage-limited power minimization in the presence of HARQ feedback.} In this scenario, the transmitter uses the HARQ feedback signals to learn about the channel condition  and update the data (re)transmission powers in a reinforcement-based fashion. As demonstrated, the proposed scheme improves the power efficiency of the HARQ protocols remarkably. For instance, consider a communication setup utilizing repetition time diversity (RTD) HARQ with codewords of rate 1 nats-per-channel-use (npcu) and outage probability $10^{-2}.$ Then, compared to uniform and the adaptive (non-reinforcement based) power allocation scheme of \cite{greenkhodemun}, the implementation of reinforcement scheme improves the power efficiency by $4$ and $1$ dB, respectively; The result is valid for a large range of fading correlations.
\end{itemize}

The remainder of the paper is organized as follows. In Section II, the system model is presented. Sections III and IV present the results for Problems 1 and 2, respectively. The conclusions are presented in Section V.

\vspace{-0mm}
\section{System model}
Consider a communication setup where, at time slot $t$, the power-limited input message $x(t)$ multiplied by the fading coefficient $h(t)$ is summed with an independent and identically distributed (iid) complex Gaussian noise $z(t) \sim \mathcal{CN}(0,1)$ resulting in the output
\vspace{-0mm}
\begin{equation}
\vspace{-0mm}
y(t) = h(t) x(t) + z(t).
\vspace{-0mm}
\end{equation}
We study temporally-correlated Rayleigh block-fading conditions where the channel coefficients remain constant in a fading block, determined by the channel coherence time, and then change to other values according to the fading probability density function (pdf). Particularly, the channel changes in each codeword transmission period according to a first-order Gauss-Markov process
\begin{align}\label{eq:channelmodelcorrelated}
h(t+1)=\beta h(t)+\sqrt{1-\beta^2}\epsilon, \epsilon\sim\mathcal{CN}(0,1).
\end{align}
Here, $\beta$ is the correlation factor of the fading realizations experienced in two successive codeword transmissions, with $\beta = 0$ (respectively, $\beta = 1$) representing the uncorrelated (respectively, fully-correlated) block-fading channel. This is a well-established model considered in the literature for different phenomena such as CSI imperfection, estimation error and channel/signals correlation \cite{5419086,cog2,heathcorrelated,846501}.
In this way, defining the channel gain as $g(t)\doteq|h(t)|^2,$ the joint and the marginal pdfs of the channel gains are found as
\begin{align}\label{eq:jointpdfr}
{f_{{g_{(t)}},{g_{(t+1)}}}}(x,y) = \frac{{{1}}}{{1 - {\beta ^2}}}{e^{ -  \frac{{x + y}}{{1 - {\beta ^2}}}}}{B_0}(\frac{{2\beta  \sqrt {xy} }}{{1 - {\beta ^2}}})
\end{align}
and
\begin{align}\label{eq:marignalpdfr}
f_{g(t)}(x)=e^{-x},x\ge 0,
\end{align}
respectively,
%and, e.g.,
%\begin{align}
%{f_{{G_{\text{ij}}}}}(x) = \lambda {e^{ - \lambda x}},x \ge 0
%\end{align}
where $B_0(.)$ is the zeroth-order modified Bessel function of the first kind \cite{846501}.

In each block, the channel coefficient is assumed to be known by the receiver, which is an acceptable assumption in block-fading channels \cite{33,noisyARQkhodemun,1715541,ekbatanioutage,6510028,Tcomkhodemun,outageHARQ,tuninetti2011,greenkhodemun,6104175,5426285}. However, there is no instantaneous channel state information available at the transmitter except the reinforcement-based feedback signals.
%
%It is assumed that there is perfect channel state information (CSI) available at the receivers.
%
Moreover, all results are presented in natural logarithm basis, the throughput is presented in npcu and the arguments are restricted to Gaussian input distributions. Finally, we concentrate on the \emph{continuous} data communication models where there is a large pool of information to be sent to the receiver, and a new codeword transmission starts as soon as the previous codeword transmission ends.

In Sections III and IV, we study Problems 1 and 2, respectively. Note that the considered problems are only examples and the reinforcement algorithms are applicable in various setups/problem formulations.
%The results are given for Rayleigh fading channels where $h \sim \mathcal{CN}(0,1)$ and, as a result, $f_g(x)=e^{-x}$ with $f_g$ denoting the channel gain pdf. In each block, the channel coefficient is assumed to be known by the receiver, which is an acceptable assumption in block-fading channels \cite{1661837,throughputdef,tuninetti2011,5336856,noisyARQkhodemun,outageHARQ,ARQGlarsson,4356994,greenkhodemun,powerarq2007,5771499,5452208,6620483,yanginft}. However, there is no instantaneous channel state information available at the transmitter except the ARQ feedback bits\footnote{The transmitter is assumed to know the long-term channel statistics, as it is required for parameter optimization.}.
%
%We consider Type-I ARQ with a maximum of $M-1$ retransmissions, i.e., the data is (re)transmitted a maximum of $M$ times, and in each round the receiver disregards the previous messages, if received in error. Also, we define a packet as the transmission of a codeword along with all its possible retransmissions. Finally, the results are obtained for the frequency-hopping based schemes where the fading coefficient changes in each retransmission independently.

\vspace{-0mm}
\section{Power-limited throughput maximization via reinforcement-based CSI feedback}
Considering the static (non-reinforcement based) CSI quantization scheme with $N$ quantization regions, an encoding function
\begin{align}\label{eq:QCSIencoding}
C(g(t))&=c_i, \,\,\text{if} \,\,g(t)\in G_i=[g_{i-1},g_i), i=1,\ldots,N,\nonumber\\& g_0\doteq0,g_N\doteq\infty,
\end{align}
is applied at the receiver and the symbol $c_i$ is fed back to the transmitter \cite{33,Tcomkhodemun}. Receiving $c_i,$ the transmitter sends the data at rate $r_i$ and power $P$.\footnote{As Section III studies the effect of reinforcement algorithms on the rate adaptation, we consider a constant (peak) power $P$. It is straightforward to extend the results to cases with adaptive power allocation.} If the instantaneous channel gain supports the data rate, i.e., $\log(1+g(t)P)\ge r_i,$ the data is successfully decoded, otherwise outage occurs. In \cite{33,Tcomkhodemun}, it has been proved that, to maximize the power-limited throughput, the optimal rate allocation rule of the static quantization schemes is given by $r_i=\log(1+\tilde g_i P)$ where
\begin{align}\label{eq:optrate}
\tilde g_i=\left\{ \begin{array}{l}
 \tilde g_1\in[0,g_1),\,\,\text{if}\,\, i=1\\
 g_{i-1}, \,\,\,\,\,\,\,\,\,\,\,\,\,\,\,\,\,\,\,\text{if}\,\, i\ne 1.\\
 \end{array} \right.
 \end{align}
That is, to maximize the throughput, the channel gain is assumed to be its worst value within each quantization region, except the first one. In this way, using $r_i=\log(1+\tilde g_i P)$ and (\ref{eq:optrate}), the throughput of the static quantized CSI scheme is determined as
\begin{align}\label{eq:etaqcsi}
\eta^\text{SQ}&=E\{\text{Achievable rates}\}\nonumber\\&=\sum_{n=1}^N{\Pr(g(t)\in G_n)\Pr(g(t)\ge r_n|g(t)\in G_n)r_n}\nonumber\\&=\sum_{n=1}^N{\log(1+\tilde g_nP)\bigg(F_{g(t)}(\tilde g^{n+1})-F_{g(t)}(\tilde g_{n})\bigg)}\nonumber\\&=\sum_{n=1}^N{\log(1+\tilde g_nP)(e^{-\tilde g_{n}}-e^{-\tilde g_{n+1}})},
 \end{align}
where $E\{.\}$ denotes the expectation operator, $F_{g(t)}(.)$ is the cumulative distribution function (cdf) of the channel gain and the last equality is for Rayleigh-fading channels. Using (\ref{eq:etaqcsi}), the power-limited throughput maximization problem of a static CSI quantization approach is formulated as
\begin{align}\label{eq:opteta}
\mathop {\max }\limits_{ \forall \tilde g_i,i=1,\ldots,N} \,\, \sum_{n=1}^N{\log(1+\tilde g_nP)(e^{-\tilde g_{n}}-e^{-\tilde g_{n+1}})}
 \end{align}
and, as the problem is complex, the optimization parameters $\tilde g_i$'s are determined via iterative optimization algorithms, e.g., \cite[Algorithm 1]{33}, \cite[Algorithm 1]{Tcomkhodemun}. Finally, setting $N=1$ and $N\to \infty,$ the throughput with no and perfect CSIT are respectively found as \cite[Chapter 1.4.1]{phdthesis}
\begin{align}
\eta^\text{No CSIT}=\mathop {\max }\limits_{ \tilde g_1} \,\, \{{e^{-\tilde g_1}\log(1+\tilde g_1P)}\}=\Lambda(P)e^{-\frac{e^{\Lambda(P)}-1}{P}}
 \end{align}
and
\begin{align}
\eta^\text{Perfect CSIT}=\int_0^\infty{e^{-g}\log(1+gP)\text{d}g}=e^{-\frac{1}{P}}\text{Ei}(-\frac{1}{P}),
 \end{align}
with $\Lambda(.)$ and $\text{Ei}(.)$ representing the Lambert W function and the exponential integral function, respectively.

Compared to no-CSIT (open-loop) systems, the static quantizers increase the throughput considerably \cite{33,Tcomkhodemun}. However, as also demonstrated in (\ref{eq:etaqcsi}), the channel temporal dependencies are not exploited for throughput increment/feedback load reduction. On the other hand, exploiting the temporal correlations has been previously shown to be crucial for practical implementation of many communication systems \cite{1715541,4389755}\footnote{For instance, the amount of CSIT required for proper implementation of
orthogonal frequency-division multiplexing (OFDM) and multiple-input-multiple-output (MIMO) broadcast
channels is not practically affordable if temporal and frequency correlations are not exploited \cite{1715541,4389755}.}.

To exploit the temporal dependencies of the channel, we propose a simple reinforcement-based algorithm as stated in Algorithm 1. In words, the algorithm is based on the following procedure. Start the data transmission with an initial transmission rate $R$ and consider an adaptation coefficient $\delta$. In each block, depending on whether the channel can support the data rate $R+\delta R$ or not, the receiver sends a reinforcement signal $\alpha=1$ or $\alpha=0,$ respectively. Receiving the reward-punishment signal $\alpha,$ the transmitter updates its transmission rate correspondingly (For more details please see Algorithm 1). The throughput is achieved by averaging on the decodable rates over many codeword transmissions.
\begin{algorithm}
\caption{CSI-based data transmission by a reinforcement algorithm}
%\begin{algorithmic}
Consider an initial transmission rate $R$ and an updating coefficient $\delta.$ In each block, do the followings.
\begin{itemize}

\item[I.] \emph{Feedback report at the receiver}

  Feed $\alpha=1$ back, if $\log(1+g(t)P)>R+\delta R.$

  Otherwise, send $\alpha=0.$
\item[II.] \emph{Rate adaptation at the transmitter}

$R\leftarrow R+\delta R,$ if $\alpha=1$

$R\leftarrow R-\delta R,$ if $\alpha=0.$

Send a codeword with rate $R$. The codeword is correctly decoded by the receiver if $\log(1+g(t)P)>R.$ Otherwise, the codeword is dropped and an outage is declared.

\item[III.] Go to I.
\end{itemize}
\end{algorithm}

As opposed to the static quantization scheme, there are only two optimization parameters in Algorithm 1 which can be determined by, e.g., exhaustive search. Also, in contrast to the static quantization scheme, the reinforcement-based scheme of Algorithm 1 \emph{follows} the channel variations and dynamically updates the transmission rates. Finally, note that to represent $N$ quantization regions $\log_2N$ feedback bits per codeword is required in  the static quantization scheme which, depending on the number of quantization regions, can be considerably high. However, the proposed algorithm is based on only 1 bit feedback per codeword.

As an example, Fig. 1 demonstrates the throughput of the reinforcement-based scheme in a Rayleigh-fading channel following (\ref{eq:channelmodelcorrelated}). Also, the results are compared with the cases having perfect and no CSIT and when the static quantization (\ref{eq:QCSIencoding}) is implemented. The parameters $R$ and $\delta$ of Algorithm 1 are found by exhaustive search such that the throughput is maximized in each signal-to-noise ratio (SNR). Also, the throughput of the static quantization scheme is obtained with $N=2$ quantization regions which leads to 1 bit per codeword feedback, the same as in the reinforcement-based scheme. Moreover, Fig. 2 shows the relative throughput gain of the proposed scheme compared to the static CSI quantization approach, i.e., $\Delta=\frac{\eta-\eta^\text{SQ}}{\eta^\text{SQ}}\%,$ where $\eta$ is the throughput achieved via the data transmission approach of Algorithm 1. As demonstrated, the system throughput is remarkably increased by implementation of the reinforcement-based algorithm. For instance, with a correlation factor of $\beta\ge0.5$ and transmission SNR of $\ge 8 \text{dB},$ the reinforcement-based scheme results in $\ge 6\%$ increase in the relative throughput. Also, the gain of the proposed scheme increases with the SNR.
%Finally, it is straightforward to prove that, with $N=2,$ the reinforcement-based scheme outperforms the static quantization, in terms of throughput, for all fading conditions. This is because the parameters of Algorithm 1 can be set as $R=\frac{1}{2}(\log(1+\tilde g_2 P)+\log(1+\tilde g_1 P))$ and $\delta=\frac{\log(1+\tilde g_2 P)-\log(1+\tilde g_1 P)}{\log(1+\tilde g_2 P)+\log(1+\tilde g_1 P)}$ which leads to the same throughput as in (\ref{eq:etaqcsi}), independently of the correlation condition.

\begin{figure}
\vspace{-0mm}
\centering
  \includegraphics[width=1\columnwidth]{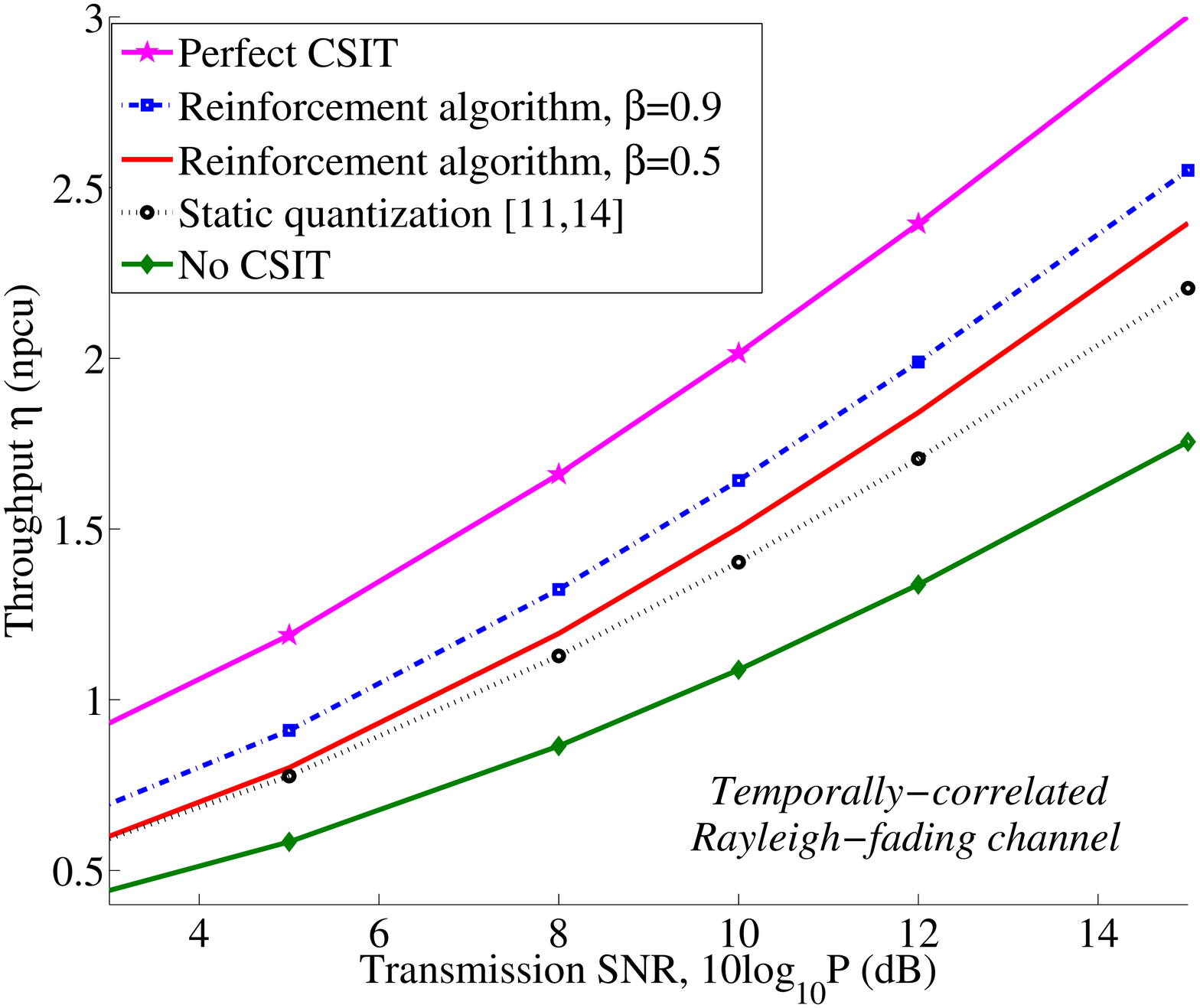}\\\vspace{-3mm}
\caption{The throughput of different schemes vs the transmission SNR $10\log_{10} P$ dB, temporally-correlated Rayleigh-fading channel following (\ref{eq:channelmodelcorrelated}).}\label{figure111}
\vspace{-0mm}
\end{figure}
\begin{figure}
\vspace{-0mm}
\centering
  \includegraphics[width=1\columnwidth]{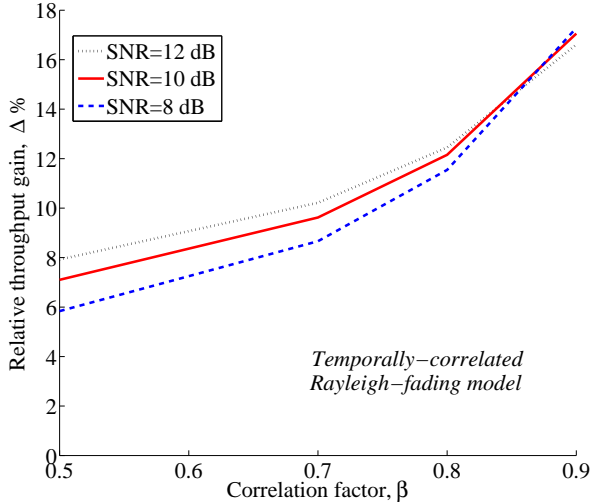}\\\vspace{-0mm}
\caption{The relative throughput gain $\Delta=\frac{\eta-\eta^\text{SQ}}{\eta^\text{SQ}}\%$ for different correlation coefficients $\beta$, temporally-correlated Rayleigh-fading channel following (\ref{eq:channelmodelcorrelated}).}\label{figure111}
\vspace{-0mm}
\end{figure}
Finally, to close the section, we should mention that, while the paper concentrates on a single-user setup, the reinforcement-based schemes are of particular interest when the number of base stations/users increases. There, the same approach as in Algorithm 1 can be implemented for user scheduling, where higher and higher data rates are considered for a user as long as it correctly decodes the message, otherwise the scheduler selects another user. Indeed, the gain of the reinforcement-based scheme, over the static quantization techniques, increases with the number of users, because to achieve the same throughput the reinforcement-based scheme requires less number of feedback bits compared to the cases with static quantization. This point becomes more interesting when we remember that, since the positive acknowledgement is a standard provision of most practical link layers \cite{6104175}, the reinforcement signal feedback is not required in each slot. As a result, the reinforcement-based scheme requires even less than one-bit feedback per user/slot.

\section{Outage-limited power minimization via reinforcement-based HARQ}
In contrast to the CSI-based schemes where the partial CSI is fed back before the codeword transmission, the HARQ-based schemes are based on reporting the message decoding status at the end of each codeword \cite{Tcomkhodemun,6104175}.

In the following, we elaborate on the implementation of reinforcement algorithms in HARQ protocols. The results are presented for the RTD HARQ, also referred to as Type III HARQ, but the discussions are valid for the other HARQ protocols, such as the incremental redundancy \cite{Tcomkhodemun,tuninetti2011}, as well.

Consider the RTD HARQ with a maximum of $M$ (re)transmission rounds, i.e., the data is retransmitted a maximum of $M-1$ times. Also, define a packet as the transmission of a codeword along with all its possible retransmissions. Using power-adaptive RTD HARQ, $b$ information nats is encoded into a codeword of length $L$ channel uses. Thus, the codeword rate is $R=\frac{b}{L}$ npcu. In the $m$th, $m=1,\ldots,M,$ (re)transmission round, the codeword is scaled to have power $P_m$. The codewords are (re)transmitted until the receiver correctly decodes the data or the maximum permitted retransmission rounds is reached. Also, in each round of a packet the receiver performs maximum ratio combining (MRC) of all received signals.

Considering the non-reinforcement based scheme and the continuous data communication model, the average power and the outage probability are obtained as follows (please see \cite{greenkhodemun} as well). In the $m$th (re)transmission round, the transmission energy is $LP_m$. If the data transmission stops at the end of the $m$th (re)transmission round, the average power, i.e., the ratio of the total transmission energy and the total data transmission time, is $P_{(m)}=\frac{\sum_{n=1}^m{LP_n}}{mL}=\frac{1}{m}\sum_{n=1}^m{P_n}.$ Thus, the average power, averaged over many packet transmissions, is obtained as
\begin{align}\label{eq:averagepow1}
P^\text{HARQ}&=\sum_{n=1}^M{(\frac{1}{m}\sum_{n=1}^m{P_n})\Pr(A_m)},
\end{align}
where $A_m$ is the event that the data transmission stops at the end of round $m.$ Note that $\sum_{m=1}^M{\Pr(A_m)}=1,$ as a maximum of $M$ (re)transmissions is considered.

As the same codeword is retransmitted, the equivalent data rate decreases to $\frac{b}{mL}=\frac{R}{m}$ at the end of the $m$th round. Also, the implementation of MRC increases the received SNR to $\sum_{n=1}^m{g(n)P_n}$ in round $m.$ Thus, following the same procedure as in \cite{tuninetti2011,greenkhodemun}, the data is successfully decoded at the end of the $m$th round (and not before) if $\log(1+\sum_{n=1}^{m-1}{g(n)P_n})<R\le\log(1+\sum_{n=1}^{m}{g(n)P_n}).$ This is based on the fact that, with an SNR $x$, the maximum achievable rate is
\begin{align}
U_{(m)}=\frac{1}{m}{\log(1+x)},\nonumber
\end{align}
if the same codeword is retransmitted $m$ times.

In this way, the probability terms $\Pr(A_m)$ are obtained as
\begin{align}\label{HARQrtd2}
\Pr(A_m)=\begin{cases}
\Pr\bigg(\log(1+\sum_{n=1}^{m-1}{g(n)P_n})<R\\\,\,\,\,\,\,\,\,\,\,\,\,\,\,\,\,\,\le\log(1+\sum_{n=1}^{m}{g(n)P_n})\bigg),  &  m\ne M \\
 1-\sum_{n=1}^{M-1}{\Pr(A_n)}, &  m=M
\end{cases}\nonumber
\\
=\begin{cases}
\Pr\bigg(\log(1+\sum_{n=1}^{m-1}{g(n)P_n})<R\\\,\,\,\,\,\,\,\,\,\,\,\,\,\,\,\,\,\le\log(1+\sum_{n=1}^{m}{g(n)P_n})\bigg),  &  m\ne M \\
 \Pr\bigg(\log(1+\sum_{n=1}^{M-1}{g(n)P_n})<R\bigg), &  m=M,
\end{cases}
\end{align}
and, with the same arguments, the outage probability is found as \cite{tuninetti2011,greenkhodemun}
\begin{align}\label{eq:HARQrtd3}
\Pr(\text{Outage})=\Pr\bigg(\log(1+\sum_{n=1}^{M}{g(n)P_n})<R\bigg).
\end{align}
Therefore, with an initial rate $R$, (\ref{eq:averagepow1})-(\ref{eq:HARQrtd3}) are used to rephrase the outage-limited power minimization problem as
\begin{align}\label{eq:outageopt}
& \mathop {\min }\limits_{ P_m,m=1,\ldots,M} \,\,\,\,\,\,\, \sum_{n=1}^M{(\frac{1}{m}\sum_{n=1}^m{P_n})\Pr(A_m)}, \nonumber\\&
 \,\,\,\,\,\,\,\,\,\,\,\,\,\text{s.t.}\,\,\,\, \,\, \Pr(\log(1+\sum_{n=1}^{M}{g(n)P_n})<R)=\epsilon,
\end{align}
where $\epsilon$ denotes the outage probability constraint. Finally, as discussed in, e.g., \cite{tuninetti2011,greenkhodemun}, there may be no closed-form solution for the optimal, in terms of (\ref{eq:outageopt}), powers $P_m$ and, depending on the fading pdf and the number of retransmissions, we may need to find the optimal power allocation rules numerically.

The drawback of power allocation based on (\ref{eq:outageopt}) is that the channel quality information gathered in the previous packet transmissions is not exploited for parameter adaptation in the next packet. That is, the power terms of a packet are not affected by the message decoding status of the previous packet transmissions. To tackle this problem, we propose a reinforcement-based algorithm as illustrated in Algorithm 2.

In the algorithm, the data transmission starts with some initial power. Then, in each time slot, depending on whether the message is correctly decoded or not, the transmission power decreases or increases, respectively. In this way, the feedback signal makes it possible to \emph{learn} about the channel quality and update the power based on all previous message decoding status. The initial power $P_\text{initial}$ and the adaptation coefficients $d_m,d'_m, m=1,\ldots,M, d_m\in(0,1),$ of the algorithm are determined by exhaustive search such that the average transmission power, averaged over many packet transmissions, is minimized for an outage probability constraint. Finally, note that to implement the reinforcement algorithm we changed the feedback model of the quantized CSI scheme in Section III. However, Algorithm 2 uses the same acknowledgement/negative acknowledgement (ACK/NACK) signal as in the standard HARQ to perform parameter adaptation.

As an example, setting $R=1$ npcu and $\beta=0.9$, Fig. 3 demonstrates the outage-limited average power of the RTD protocol with a maximum of $M=2$ (re)transmissions. Also, the results are compared with the cases utilizing uniform power allocation, i.e., $P_m=P_n,\forall m,n,$ and when the power terms are optimized based on (\ref{eq:outageopt}). To solve (\ref{eq:outageopt}), we have used the same iterative optimization algorithm as in \cite[Algorithm 1]{greenkhodemun}. As it can be seen, remarkable power efficiency gain is achieved by the reinforcement algorithm. For instance, with an outage probability $\epsilon=10^{-2},$ the implementation of the reinforcement-based algorithm reduces the average power, compared to the uniform power allocation and the power allocation scheme of (\ref{eq:outageopt}), by $4$ and $1$ dB, respectively. Also, the effect of reinforcement algorithm increases as the outage probability constraint becomes harder, i.e., $\epsilon$ decreases. Finally, although not demonstrated in the figure, (almost) the same average power reduction is observed for the cases with $\beta\ge 0.2.$

\begin{figure}
\vspace{-0mm}
\centering
  \includegraphics[width=1\columnwidth]{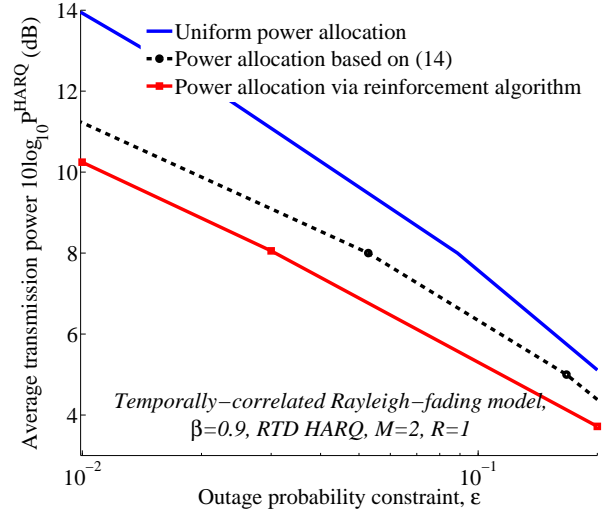}\\\vspace{-0mm}
\caption{Outage-limited average power for different power allocation schemes, RTD HARQ, $M=2$. Correlated Rayleigh fading channel model (\ref{eq:channelmodelcorrelated}), $\beta=0.9$. For the reinforcement-based scheme, Algorithm 2 is used where the constants $P_\text{initial},d_m,d'_m,\forall m,$ are optimized, in terms of average power, for every given outage probability.  }\label{figure111}
\vspace{-0mm}
\end{figure}

\begin{algorithm} [tbh]
\caption{HARQ-based data transmission by a reinforcement algorithm}
%\begin{algorithmic}
\begin{itemize}
\item[I.] For a given initial transmission rate $R$, set the initial transmission power to $\breve{P}=P_\text{initial}$ and consider the adaptation coefficients $d_m,d'_m,m=1,\ldots,M,d_m\in (0,1).$
\item[II.] Start a new packet transmission with power $\breve{P}$ and do the following procedure
\begin{itemize}
  \item[1)] For $m<M$,

  If the codeword is correctly decoded, set $\breve{P}\leftarrow(1-d_m)\breve{P},$ $m\leftarrow 1$ and go to II.

  If the codeword is not decoded, set $\breve{P}\leftarrow(1+d'_m)\breve{P},$ $m\leftarrow m+1$ and retransmit the codeword.

  \item[2)]  For $m=M$,

  If the codeword is correctly decoded, set $\breve{P}\leftarrow(1-d_M)\breve{P},$ $m\leftarrow 1$ and go to II.

  If the codeword is not decoded, declare an outage, set $\breve{P}\leftarrow(1+d'_M)\breve{P},$ $m\leftarrow 1$ and go to II.
\end{itemize}

\end{itemize}

\end{algorithm}

%\vspace{-2mm}

\section{Conclusion}
This paper studied the data transmission efficiency of the communication systems utilizing reinforcement algorithms. Considering temporally-correlated fading channels, the reinforcement feedback signals were used for parameter adaptation in the cases with partial CSIT. As illustrated, the reinforcement algorithms lead to remarkable performance improvement, compared to the state-of-the-art schemes, with the same feedback load. Specially, considerable throughput and power efficiency increment is achieved with 1 bit per codeword feedback, if the reinforcement algorithms are utilized.
%\section*{Acknowledgement}
%This work was supported in part by the Swedish Governmental Agency for Innovation Systems (VINNOVA) within the VINN Excellence Center Chase.
\vspace{-0mm}
\bibliographystyle{IEEEtran} %lic.bst is the style file
\bibliography{masterfiniteblock}
\vfill
\end{document}